\documentclass[prb,twocolumn,showpacs,preprintnumbers,amsmath,amssymb]{revtex4}
\usepackage{graphicx}

\begin{document}

\title{Exact diagonalization study of optical conductivity in two-dimensional Hubbard model}

\author{T. Tohyama}
\email{tohyama@imr.tohoku.ac.jp}
\author{Y. Inoue}
\author{K. Tsutsui}
\author{S. Maekawa}
 \altaffiliation[Also at ]{CREST, Japan Science and Technology Agency (JST), Kawaguchi, 
Saitama 332-0012, Japan.}
\affiliation{Institute for Materials Research, Tohoku University, Sendai, 980-8577, Japan.}
\date{\today}

\begin{abstract}
The optical conductivity $\sigma(\omega)$ in the two-dimensional Hubbard model is examined by applying the exact diagonalization technique to small square clusters with periodic boundary conditions up to $\sqrt{20}\times\sqrt{20}$ sites.  Spectral-weight distributions at half filling and their doping dependence in the 20-site cluster are found to be similar to those in a $\sqrt{18}\times\sqrt{18}$ cluster, but different from $4\times 4$ results.  The results for the 20-site cluster enable us to perform a systematic study of the doping dependence of the spectral-weight transfer from the region of the Mott-gap excitation to lower-energy regions.  We discuss the dependence of the Drude weight and the effective carrier number on the electron density at a large on-site Coulomb interaction.
\end{abstract}


\maketitle

\section{Introduction}
Charge carriers introduced into the Mott insulators induce a dramatic change of the electronic states. The high-temperature cuprate superconductor is a good example, where the spectral weight of the Mott-gap excitation around $\sim2$~eV in the optical conductivity, $\sigma(\omega)$, decreases rapidly upon carrier doping and new weight emerges in the mid-infrared region accompanied with the increase of a zero-frequency component that corresponds to the dc conductivity.~\cite{Uchida,Arima,Onose,Waku}  Such a reconstruction of the electronic states would not occur in a conventional semiconductor.

The electronic states of the cuprates are known to be well described by a single-band Hubbard model with a large on-site Coulomb interaction in two dimensions.  Numerous theoretical works have been done to understand the physics of the two-dimensional (2D) Hubbard model.  Among them, the numerical diagonalization technique based on the Lanczos algorithms has been intensively used for the investigation of the rapid spectral weight transfer seen in $\sigma(\omega)$ and other quantities.~\cite{Dagotto,Nakano,Eskes,Becca,Bonca}  The doping dependence of $\sigma(\omega)$ for a $4\times4$ cluster under periodic boundary conditions has clearly shown the reconstruction of the weight.~\cite{Dagotto}  However, the cluster significantly suffers from the finite-size effect.  For example, the Drude weight $D$ at half-filling, where the electron density $n=1$ (16 electrons in the cluster), becomes negative, and the ground state at $n=0.875$ (14 electrons) is unphysically threefold degenerated.  Although $D$ seems to grow proportional to $1-n$ near half filling,~\cite{Dagotto} this would not be conclusive because of these finite-size effects.  It has been argued~\cite{Nakano} that $D$ in a $4\times 4$ cluster with mixed boundary conditions rather grows proportional to $(1-n)^2$.  In these situations, it is necessary to perform a systematic study of the optical conductivity by increasing the size of clusters.

In this paper, we perform the exact diagonalization study of $\sigma(\omega)$ in the 2D Hubbard model up to a 20-site cluster.  The diagonalization of the 20-site cluster has not been done in the past as far as we know. 
 We compare $\sigma(\omega)$ for $4\times 4$, $\sqrt{18}\times\sqrt{18}$, and $\sqrt{20}\times\sqrt{20}$ clusters with periodic boundary conditions.  We find that at half-filling the 18- and 20-site results are similar in spectral shape but different from the 16-site result about the spectral weight distribution. The doping dependence of $\sigma(\omega)$ also shows similar behavior between the 18- and 20-site lattices.  For the 20-site cluster, we perform a systematic study of the doping dependence of the spectral-weight transfer, and find that the transferred weight from the Mott-gap excitation increases with doping, showing a behavior qualitatively consistent with that observed experimentally.~\cite{Uchida} We also find that $D$ grows roughly proportional to $1-n$ near half-filling.  These exact results can be useful in understanding the physics of the 2D Hubbard model. 

This paper is organized as follows.  In Sec.~\ref{Model}, we introduce the 2D Hubbard model and the method for the exact diagonalization.  The doping dependence of the ground-state energy is also presented in the section. In Sec.~\ref{Optical}, the calculated results of $\sigma(\omega)$ are shown.  The doping dependence of the spectral-weight transfer, as well as of the Drude weight, is performed for the 20-site cluster.  The summary is given in  Sec.~\ref{Summary}.

\section{Model and method}
\label{Model}
The Hamiltonian of the Hubbard  model in 2D reads as
\begin{equation}
H= -t\sum_{\mathbf{i},\boldsymbol{\delta},\sigma}
    \left(c_{\mathbf{i}+\boldsymbol{\delta},\sigma }^\dagger c_{\mathbf{i},\sigma}+c_{\mathbf{i}-\boldsymbol{\delta},\sigma }^\dagger c_{\mathbf{i},\sigma} \right) + U\sum_\mathbf{i} n_{\mathbf{i},\uparrow} n_{\mathbf{i},\downarrow},
\end{equation}
where $t$ is the hopping energy, $U$ is the on-site Coulomb interation, and $\boldsymbol{\delta}=\mathbf{x}$ and $\mathbf{y}$, $\mathbf{x}$ and $\mathbf{y}$ are the vectors connecting the nearest-neighbor sites in the $x$ and $y$ directions, respectively.  The operator $c_{\mathbf{i},\sigma}^\dagger$ ($c_{\mathbf{i},\sigma}$) creates (annihilates) an electron with spin $\sigma$ at site $\mathbf{i}$, and $n_{\mathbf{i},\sigma}=c_{\mathbf{i},\sigma}^\dagger c_{\mathbf{i},\sigma}$.

We apply the exact diagonalization method based on the Lanczos technique to square clusters with periodic boundary conditions up to 20 sites.  The size of the Hilbert space  to be diagonalized can be reduced by employing the translational and point-group symmetries of the clusters.  In a $4\times 4$ cluster, the maximum size of the Hamiltonian matrix is 5,178,144 for the space with $n=1$ (half-filling), null $z$-component of the total spin ($S_z=0$), the total momentum $\mathbf{K}=(\pm\pi/2,0)$, $(0,\pm\pi/2)$, $(\pi,\pm\pi/2)$, and $(\pm\pi/2,\pi)$ with A$_\mathrm{1}$ irreducible representation.  Since the memory to be required for a single vector is 79~MB (complex with double precision), a standard workstation (Pentium 4,  2.67~GHz,  1~GB memory) is enough for the Lanczos diagonalization: the ground-state energy is obtained in 4~h by using our code.  The size of the matrix rapidly increases with the size of cluster. In a $\sqrt{20}\times\sqrt{20}$ cluster, the maximum size is 1,706,748,500 for the space with  $n=1$, $S_z=0$, $\mathbf{K}=(\pm 4\pi/5,\pm 2\pi/5)$ and $(\pm 2\pi/5,\mp 4\pi/5)$ (25.43~GB for a single vector). For such a huge matrix, we made use of the supercomputer (Hitachi SR11000 in the Institute for Molecular Science, Okazaki) and a parallelization based on the message passing interface.  In order  to obtain the ground-state energy, it takes 37~h when we use 16~nodes in the supercomputer.  

\begin{figure}
\begin{center}
\includegraphics[width=7.5cm]{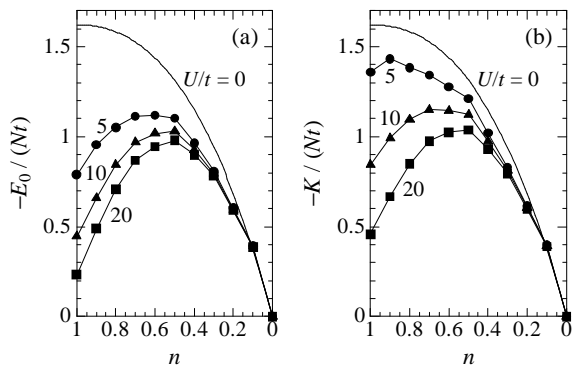}
\caption{\label{energy}
Ground-state energy $E_0$ (a) and energy of the hopping part (kinetic energy) $K$ (b) per site for the $\sqrt{20}\times\sqrt{20}$ cluster ($N=20$) with an even number of electrons.  $U/t=5$ (circles), 10 (upper triangles), and 20 (squares). The energies for $U/t=0$ (solid lines) are obtained from the infinite-size system.}
\end{center}
\end{figure}

The ground-state energy $E_0$ of the $\sqrt{20}\times\sqrt{20}$ cluster with an even number of electrons is plotted in Fig.~\ref{energy}(a) as a function of $n$ for $U/t=5$, 10, and 20.  In the weak-coupling regime with a small value of $U$, the ground-state energy can be obtained by ordinary perturbation theory.~\cite{Metzner}. For $U/t\ge$5 we find no indication of phase separation from Fig.~\ref{energy}(a).
We also find that the energy curve does not behave smoothly at $n=0.5$.  This deviation probably comes from a closed-shell structure formed in the momentum space when $n=0.5$ (ten electrons in the cluster).  In other words, the kinetic-energy term may be responsible for the deviation. In fact, the same deviation is seen in the kinetic energy, as shown in Fig.~\ref{energy}(b).

As a database, we have listed in Ref.~11 
the ground-state energies of the $\sqrt{18}\times\sqrt{18}$ and $\sqrt{20}\times\sqrt{20}$ clusters with $U/t=5$, 10, and 20 for all $n$, together with the lowest eigenenergy in each subspace of the clusters. We note that for $U/t=5$, 10, and 20 the ground states of the $\sqrt{20}\times\sqrt{20}$ cluster with an even number of electrons are always located in the subspace of zero total momentum, but not for $\sqrt{18}\times\sqrt{18}$. 

\section{Optical conductivity}
\label{Optical}
The real part of the optical conductivity under the electric field applied along the $x$ direction is given by
$\sigma(\omega)=2\pi D \delta(\omega)+\sigma_\mathrm{reg}(\omega)$,
where $D$ is the charge stiffness, which is sometimes called the Drude weight. The regular part $\sigma_\mathrm{reg}(\omega)$ is 
\begin{equation}
\sigma_\mathrm{reg}(\omega)=\frac{\pi}{N\omega} \sum_{m\neq 0} \left|\left< \Psi_m \left| j_x \right|\Psi_0\right>\right|^2 \delta (\omega- E_m + E_0)\;,
\label{Regular}
\end{equation}
where $N$ is the number of sites, $\Psi_0$ ($\Psi_m$) is the ground (final) state with energy $E_0$ ($E_m$), and the electric charge $e$, the lattice spacing $a$, and $\hbar$ are taken to be unity.  
The $x$ component of the current operator is given by
$j_x=-i t \sum_{\mathbf{i},\sigma}
    ( c_{\mathbf{i}+\mathbf{x},\sigma }^\dagger c_{\mathbf{i},\sigma}-c_{\mathbf{i}-\mathbf{x},\sigma }^\dagger c_{\mathbf{i},\sigma} )$.
From the $f$-sum rule, $D$ is given by
\begin{equation}
D=-\frac{1}{2N}\left< \Psi_0 \left| \tau_{xx} \right|\Psi_0\right>-\frac{1}{\pi}\int_0^\infty\sigma_\mathrm{reg}(\omega) d\omega\;,
\label{D}
\end{equation}
where the stress-tensor operator $\tau_{xx}$ is given by
$\tau_{xx}= -t \sum_{\mathbf{i},\sigma}
    ( c_{\mathbf{i}+\mathbf{x},\sigma }^\dagger c_{\mathbf{i},\sigma}+c_{\mathbf{i}-\mathbf{x},\sigma }^\dagger c_{\mathbf{i},\sigma} )$.
We note that the double of the first term on the right-hand side of Eq.~(\ref{D}) is plotted in Fig.~\ref{energy}(b) for the $\sqrt{20}\times\sqrt{20}$ cluster.  
A standard Lanczos technique with a Lorentzian broadening of $0.2t$ is used in calculating Eq.~(\ref{Regular}).

\begin{figure}
\begin{center}
\includegraphics[width=7.5cm]{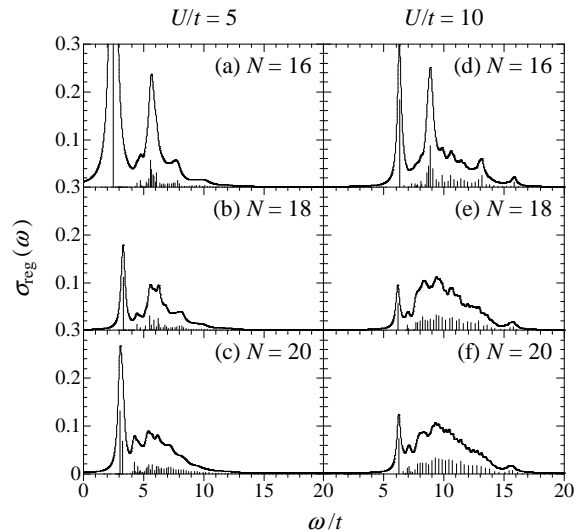}
\caption{\label{sigma_n=1}
Regular part of the optical conductivity $\sigma_\mathrm{reg}(\omega)$ at half-filling ($n=1$).  (a)-(c) $U/t=5$.  (d)-(f) $U/t=10$. (a) and (d) $4\times 4$. (b) and (e) $\sqrt{18}\times\sqrt{18}$. (c) and (f) $\sqrt{20}\times\sqrt{20}$.}
\end{center}
\end{figure}

Figure~\ref{sigma_n=1} shows the dependence of $\sigma_\mathrm{reg}(\omega)$ on the size of clusters at half-filling with $U/t=5$ and 10.  Irrespective of the size, there is a sharp peak at the Mott-gap edge.  This is an excitonic peak caused by the spin-polaron formation in the photoexcited state.~\cite{Tohyama1,Onodera}
We note that such an excitonic peak is not seen in $\sigma_\mathrm{reg}(\omega)$ of the 1D Hubbard model at half-filling.~\cite{Jeckelmann}
We find in the figure that the weight of the peak is dependent on the system size: The weights for $\sqrt{18}\times\sqrt{18}$ and $\sqrt{20}\times\sqrt{20}$ are smaller than that for $4\times 4$. Size dependence is also seen at a higher-energy region, where broad spectral weights spread: The $4\times 4$ result exhibits a pronounced peak structure, for example, at $\omega/t=9$ for $U/t=10$, which is absent in the $\sqrt{18}\times\sqrt{18}$ and $\sqrt{20}\times\sqrt{20}$ cases.  From the figure, we can conclude that the $\sqrt{18}\times\sqrt{18}$ and $\sqrt{20}\times\sqrt{20}$ results are qualitatively the same as that of $4\times 4$, but quantitatively different.  Such a quantitative difference may partly come from a special shape of the $4\times 4$ periodic cluster that possesses the symmetry of a hypercube.~\cite{Dagotto}

\begin{figure}
\begin{center}
\includegraphics[width=7.5cm]{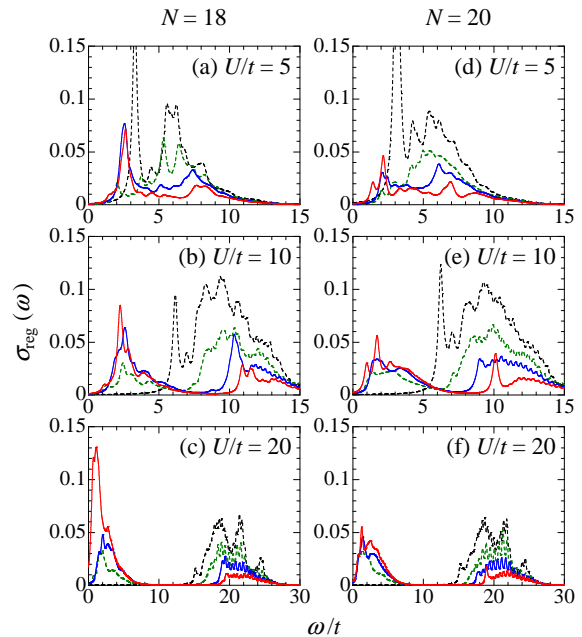}
\caption{\label{sigma_doping}
(Color online) Dependence of the regular part of the optical conductivity $\sigma_\mathrm{reg}(\omega)$ on the electron density $n$ near half-filling.  (a)-(c) $\sqrt{18}\times\sqrt{18}$. (d)-(f) $\sqrt{20}\times\sqrt{20}$.  (a) and (d) $U/t=5$.  (b) and (e) $U/t=10$. (c) and (e) $U/t=20$.  The black (dash-dotted), green (gray dash-dotted), blue (solid), and red (dark-gray solid) lines in the 20-site (18-site) cluster correspond to the results of $n=1$ (1), 0.9 (0.889), 0.8 (0.778), and 0.7 (0.667), respectively.}
\end{center}
\end{figure}

The doping dependence of $\sigma_\mathrm{reg}(\omega)$ near half-filling for $\sqrt{18}\times\sqrt{18}$ and $\sqrt{20}\times\sqrt{20}$ are shown in Fig.~\ref{sigma_doping}.  For all values of $U/t$ shown in the figure, the weight of the Mott-gap excitation decreases by decreasing the density $n$ from $n=1$, being consistent with the previous $4\times 4$ results.~\cite{Dagotto,Nakano}  The most pronounced feature is the disappearance of the excitonic peak seen at half-filling.  This is understood as a result of the suppression of antiferromagnetic order, since the excitonic-peak formation is closely related to the spin correlation in the spin background.~\cite{Tohyama1,Onodera}  The suppression of the peak with doping has recently been observed in Na-doped Ca$_2$CuO$_2$Cl$_2$.~\cite{Waku}

In contrast to the suppression of the Mott-gap excitation, the spectral weight emerges inside the gap with carrier doping.  The weight gradually increases with decreasing $n$ in both the $\sqrt{18}\times\sqrt{18}$ and $\sqrt{20}\times\sqrt{20}$ clusters.  This is clearly seen in the cases of $U/t=10$ and 20 because of an obvious separation between the in-gap and Mott-gap excitations.  For $U/t=5$, the separation is not clear, however at around $\omega/t=2$ the weight increases with decreasing $n$.  We note that in the 1D Hubbard model, no spectral weight emerges inside the gap upon carrier doping.~\cite{Fye} 

In a $4\times 4$ cluster with periodic boundary conditions,~\cite{Dagotto} the weight inside the gap was reported to increase with reducing $n$, but the result was obtained by subtracting a lowest-energy peak with large weight that was regarded as a part of the Drude weight.  On the other hand, the results for a $4\times 4$ cluster with mixed boundary conditions~\cite{Nakano} showed a decrease of the in-gap weight with decreasing $n$.  This contradictory behavior in the two $4\times 4$ clusters seems to come from the finite-size effect that is enhanced in the case of 14 electrons in the clusters.  In our $\sqrt{18}\times\sqrt{18}$ and $\sqrt{20}\times\sqrt{20}$ clusters, however, the doping dependence of the in-gap weight near half-filling shows the same tendency and is consistent with the experimental data.~\cite{Uchida,Arima,Onose,Waku}  We note that a dramatic increase of the low-energy peak at $n=12/18\sim0.667$ in Fig.~\ref{sigma_doping}(c) is probably due to nonzero total momentum of the ground state [$\mathbf{K}=(\pm 2\pi/3,\pm 2\pi/3)$ (Ref.~11)].

\begin{figure}
\begin{center}
\includegraphics[width=6.5cm]{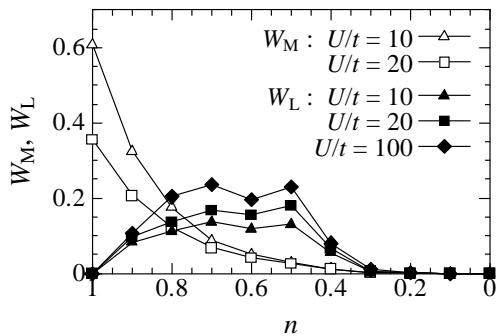}
\caption{\label{weight}
Dependence of integrated spectral weights in $\sigma_\mathrm{reg}(\omega)$ on the electron density $n$ for the $\sqrt{20}\times\sqrt{20}$ cluster.  The weights of the Mott-gap excitation, $W_\mathrm{M}$, are shown by the open upper triangles and squares for $U/t=10$ and 20, respectively.   The solid upper triangles, squares, and diamonds represent the low-energy weight below the Mott gap, $W_\mathrm{L}$, for $U/t=10$, 20, and 100, respectively.}
\end{center}
\end{figure}

In order to see the spectral-weight transfer more clearly, we plot in Fig.~\ref{weight} integrated weights above and below the Mott gap for the $\sqrt{20}\times\sqrt{20}$ cluster.  The integrated weight above the Mott gap, $W_\mathrm{M}$, is obtained by subtracting the integrated weight below the Mott-gap excitation, $W_\mathrm{L}$, from the total weight of $\sigma_\mathrm{reg}(\omega)$.  The $W_\mathrm{M}$ decreases rapidly by decreasing $n$ from half-filling, and almost disappears below $n=0.3$.  This behavior is consistent with that of an analytical expression that has been derived from the large $U$ limit of the Hubbard model,~\cite{Eskes} where the nearest-neighbor spin-spin correlation as well as three-site hopping processes of electrons predominantly determines $W_\mathrm{M}$.  The $W_\mathrm{L}$, on the other hand, increases with decreasing $n$ from $n=1$, and shows two maxima at around $n=0.7$ and 0.5.  The latter maximum might be related to the closed-shell structure in the case of ten electrons, as mentioned above.  Regarding the $n=0.5$ maximum as a consequence of the finite-size effect, we find that the doping dependence of $W_\mathrm{L}$ is qualitatively similar to that in the infinite-dimensional Hubbard model,~\cite{Jarrell} where $W_\mathrm{L}$ show a broad peak at around $n=0.75$.

The $W_\mathrm{L}$ is enhanced with increasing $U$, i.e., with decreasing the AF exchange constant $J=4t^2/U$, as seen in Fig.~\ref{weight}.  This enhancement is accompanied by a reduction of $D$, as will be shown below.  The decrease of $J$ may induce the suppression of coherent motion of spin-polaronic particles but enhance their incoherent motion.~\cite{Rice}  This results in the enhancement of $W_\mathrm{L}$ together with the reduction of $D$.

\begin{figure}
\begin{center}
\includegraphics[width=7.5cm]{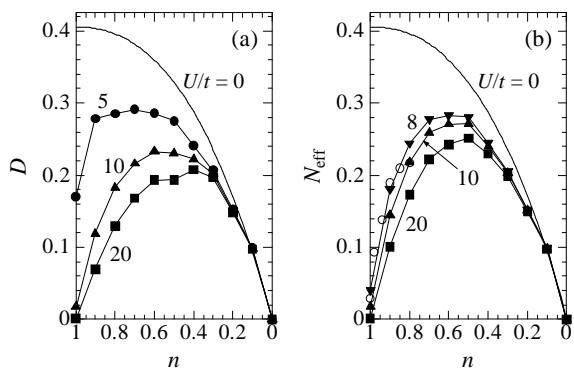}
\caption{\label{Drude+Neff}
Dependence of the Drude weight $D$ (a) and effective carrier number $N_\mathrm{eff}$ below the Mott gap (b) on the electron density $n$ for several values of $U/t$ in the $\sqrt{20}\times\sqrt{20}$ cluster.  $U/t=5$ (circles), 8 (lower triangles), 10 (upper triangles), and 20 (squares). The solid line is for $U/t=0$.  The open circles in (b) correspond to the experimental data for La$_{2-x}$Sr$_x$CuO$_4$ (Ref.~1).}
\end{center}
\end{figure}

Figure~\ref{Drude+Neff}(a) shows the $n$ dependence of $D$ for the 20-site cluster.  For large $U/t$, $D$ at half-filling ($n=1$) is close to zero, in contrast to the case of the periodic $4\times 4$ cluster, where $D$ is negative at $n=1$.  Therefore, the finite-size effect on $D$ is remarkably reduced.  However,  in $U/t=5$, $D$ at $n=1$ is finite with large value, being different from a plausible view of an insulating ground state at half-filling.~\cite{Furukawa,Gebhard}  This implies that the finite-size effect remains large for a small value of $U$.    For very large $U$, $D$ exhibits a dome-like shape as a function of $n$.  The maximum position of $D$ shifts toward $n=1$ with decreasing $U$, as expected.

The doping dependence of $D$ near half-filling would be related to the nature of the metal-insulator transition.~\cite{Imada} It has been argued~\cite{Nakano} that, at $U/t=16$, $D$ in the $4\times 4$ cluster with mixed boundary conditions grows proportional to $(1-n)^2$ near half-filling by examining the values of $D$ at $n=1$, 0.875, and 0.75. In contrast to this, our data for large $U$ in Fig.~\ref{Drude+Neff}(a) grow roughly proportional to $1-n$, with a slight convex behavior from $n=1$ to 0.8., i.e., $D\propto (1-n)^\alpha$ with $\alpha\leqslant 1$.  The results of the $\sqrt{18}\times\sqrt{18}$ cluster show the same behavior (not shown).  In the strong coupling limit of the 20-site Hubbard model (the $t$-$J$ model plus three-site terms), the tendency is unchanged under mixed boundary conditions.~\cite{Tohyama2}  A similar dependence has also been reported in a dynamical mean-field theory to incommensurate magnetic ordered states in the 2D Hubbard model.~\cite{Fleck}  However, since our clusters are still small, we need to work on larger systems to obtain more conclusive results about the $n$ dependence of $D$ near the metal-insulator transition.   

In Fig.~\ref{Drude+Neff}(b), we plot the effective carrier density $N_\mathrm{eff}$ for the $\sqrt{20}\times\sqrt{20}$ cluster, defined as~\cite{Dagotto}
\begin{equation}
N_\mathrm{eff}=\frac{1}{\pi t}\left( \pi D+\int_{0^+}^{\omega_\mathrm{c}} \sigma_\mathrm{reg}(\omega) d\omega\right)\;.
\label{Neff}
\end{equation}
where $\omega_\mathrm{c}$ is an energy just below the Mott-gap excitation.  The dependence of $N_\mathrm{eff}$ on $n$ shows a behavior qualitatively similar to that observed experimentally for La$_{2-x}$Sr$_x$CuO$_4$:~\cite{Uchida} $N_\mathrm{eff}$ has a tendency to be saturated at large doping.  The experimental data seem to be fitted well to the $U/t=8$ results, but for a more complete comparison we need to include, for example, the long-range hoppings that are necessary for the study of the material dependence.~\cite{Tohyama3} 

\section{Summary}
\label{Summary}
In summary, we have examined $\sigma(\omega)$ in the 2D Hubbard model by applying the exact diagonalization technique to small square clusters up to 20 sites under periodic boundary conditions.  The 20 sites in 2D are the largest size treated ever by the exact diagonalization as far as we know. We have found that the spectral shapes of $\sigma(\omega)$ at half-filling and their doping dependence in the 20-site cluster are similar to those in the 18-site cluster, but different from the 16-site results. By using the 20-site results, we have analyzed the doping dependence of the spectral-weight transfer as well as of $D$, and clarified systematic changes of $\sigma(\omega)$ in the 2D Hubbard model, which provide important information on the physics of the model.

\section*{Acknowledgements}
We would like to thank N. Bulut, G. Seibold, and G. Baskaran for useful discussions. This work was supported by NAREGI Nanoscience Project, CREST, and Grant-in-Aid for Scientific Research from the Ministry of Education, Culture, Sports, Science, and Technology of Japan.  The numerical calculations were partly performed in the supercomputing facilities in Tohoku University and ISSP, University of Tokyo.

\end{document}